\documentclass[sigconf,authorversion,nonacm]{acmart}
\AtBeginDocument{%
  \providecommand\BibTeX{{%
    \normalfont B\kern-0.5em{\scshape i\kern-0.25em b}\kern-0.8em\TeX}}}

\acmConference[Woodstock '18]{Woodstock '18: ACM Symposium on Neural Gaze Detection}{June 03--05, 2018}{Woodstock, NY}
\acmBooktitle{Woodstock '18: ACM Symposium on Neural Gaze Detection,June 03--05, 2018, Woodstock, NY}
\renewcommand\footnotetextcopyrightpermission[1]{}

\begin{document}

%%
%% The "title" command has an optional parameter,
%% allowing the author to define a "short title" to be used in page headers.
\title[A Human-Powered Public Display that Nudges Social Biking via Motion Gesturing]{A Human-Powered Public Display that Nudges Social Biking via Motion Gesturing}

%%
%% The "author" command and its associated commands are used to define
%% the authors and their affiliations.
%% Of note is the shared affiliation of the first two authors, and the
%% "authornote" and "authornotemark" commands
%% used to denote shared contribution to the research.
\author{Binh Vinh Duc Nguyen}
\email{alex.nguyen@kuleuven.be}
\orcid{0000-0001-5026-474X}
\affiliation{
  \institution{Research[x]Design, \break Department of Architecture, KU Leuven}
  \streetaddress{Kasteelpark Arenberg 1 - box 2431}
  \city{Leuven}
  \country{Belgium}
  \postcode{3001}
}

\author{Andrew Vande Moere}
\email{andrew.vandemoere@kuleuven.be}
\orcid{0000-0002-0085-4941}
\affiliation{
  \institution{Research[x]Design, \break Department of Architecture, KU Leuven}
  \streetaddress{Kasteelpark Arenberg 1 - box 2431}
  \city{Leuven}
  \country{Belgium}
  \postcode{3001}
}

%%
%% By default, the full list of authors will be used in the page
%% headers. Often, this list is too long, and will overlap
%% other information printed in the page headers. This command allows
%% the author to define a more concise list
%% of authors' names for this purpose.
%\renewcommand{\shortauthors}{Trovato and Tobin, et al.}

%%
%% The abstract is a short summary of the work to be presented in the
%% article.
\begin{abstract}
The WeWatt bike serves as an energy station that enables passers-by to charge their mobile devices through physical activity. However, despite multiple people using it simultaneously, the bike is typically used individually. To address this limitation, we developed the WeWattTree, an installation utilising human-powered energy to filter environmental air. Through the orchestration of subtle motion gestures, our goal is to entice passers-by to participate and encourage them to socially interact, synchronising their pace. In this work-in-progress, we provide insights into the prototyping process, combining physical experimentation and computational simulation, and delve into the underlying concepts of our grammar of motion gestures. We highlight how a single design effectively merged multiple functionalities, how the role of material characteristics shaped the interaction design, and discuss the potential for social performances as captivating public displays.
\end{abstract}

%%
%% The code below is generated by the tool at http://dl.acm.org/ccs.cfm.
%% Please copy and paste the code instead of the example below.
%%

%%
%% Keywords. The author(s) should pick words that accurately describe
%% the work being presented. Separate the keywords with commas.

%% A "teaser" image appears between the author and affiliation
%% information and the body of the document, and typically spans the
%% page.
\begin{teaserfigure}
  \centering
  \includegraphics[width=0.5\linewidth]{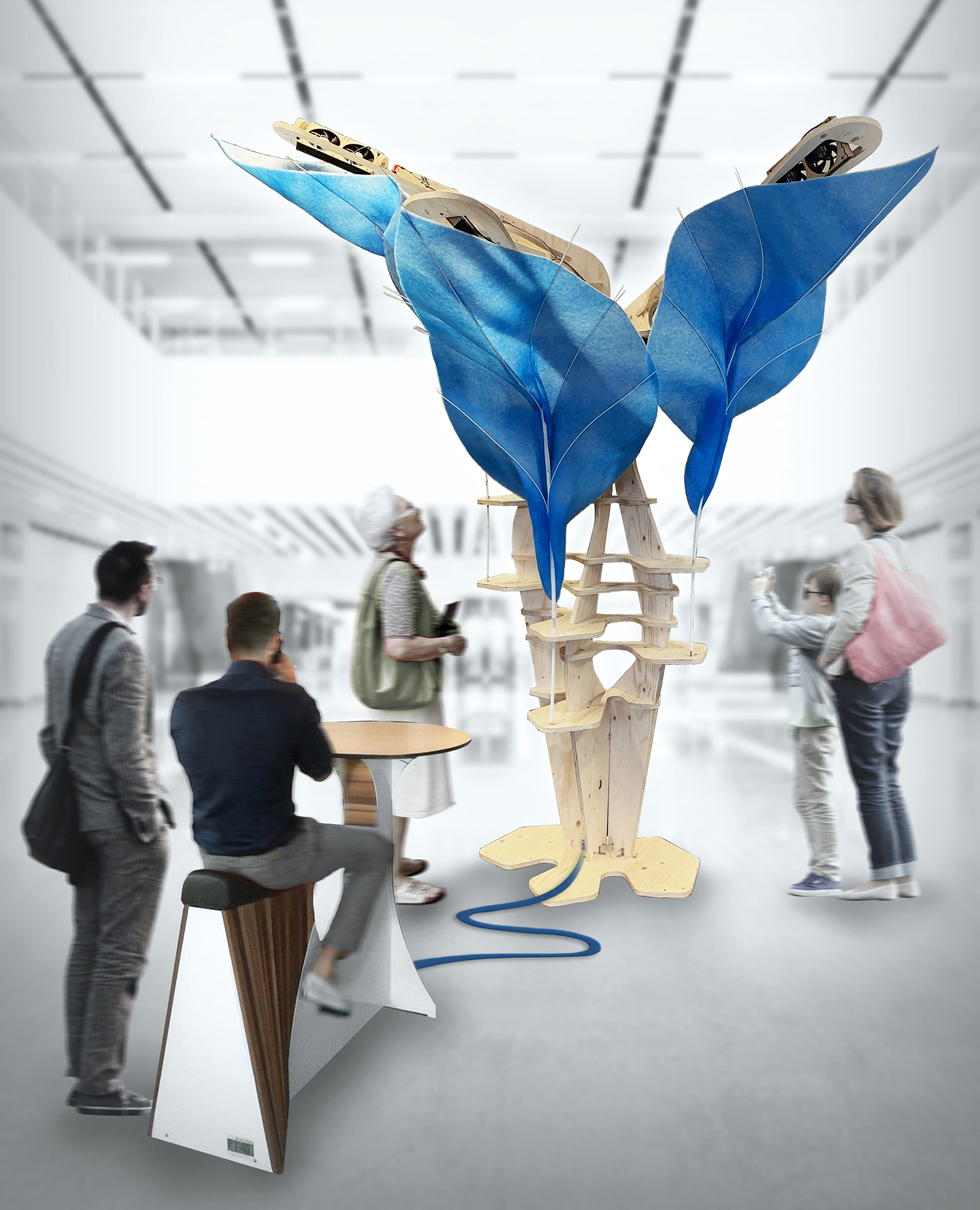}
  \caption{The WeWattTree prototype turns biking behaviour into a shared type of social engagement by providing a collective experience for bikers and bystanders.}
  \label{fig:DIS2020-tree}
\end{teaserfigure}

%%
%% This command processes the author and affiliation and title
%% information and builds the first part of the formatted document.
\maketitle

\section{Introduction}
The WeWatt bike, developed and produced by WeWatt, a Belgian company, is an energy station that allows passers-by to charge their mobile devices through their own physical effort. This innovative concept aims to promote physical activity while generating green energy. While the bike is generally well-received, our observations revealed that users tended to avoid interacting with each other while using it. To foster a sense of community, we conceived the `WeWattTree', a kinetic installation illustrated in \autoref{fig:DIS2020-tree}. Despite being technically limited to human power, the design rationale behind the WeWattTree encompassed three key goals: 1) motivating passers-by to engage with biking for a substantial period; 2) encouraging social interactions among multiple bikers; and 3) aligning all these activities with a "good" and sustainable cause.

\section{Design Rationale}
Public display technology was an apparent choice to engage passers-by initially. However, public displays face various challenges in terms of weather conditions, social events, deployment space, inhabitants, and vandalism \cite{Makela2017}. Additionally, large public displays require a large amount of power, often exceeding what can be consistently generated by human effort alone, typically ranging from \(40\) to \(60\) Watts. In contrast, there is considerable potential in exploring display-less metaphors to create more immersive and shared sensory experiences around data \cite{Hogan2016, VandeMoere2008, Nguyen2023}. These alternatives often require much less power compared to traditional light-emitting pixels.

As WeWatt required the integration of a "good" and preferably "green" cause, we saw an opportunity in combining interpretable motion gestures with the act of filtering air. This concept materialised as a set of rotating fans that convert human pedalling power into a gentle breeze, blowing air through various air purifying surfaces that also move and wave in response to the generated airflow. Previous studies showed that wind itself served as a form of display, enhancing the interactive experience through haptic modality \cite{Sawada2007}. However, precise control of wind speed, distance, and orientation was crucial to generate recognisable patterns for users \cite{Mowafi2015}. Since people perceived wind in diverse ways, the richness of a wind-based experience did not necessarily depend on the complexity of carefully designed airflow \cite{Tolley2019}. To address this, we explored the integration of alternative modalities, such as hearing \cite{Ranasinghe2017, Nguyen2023} or seeing \cite{Sodhi2013}. For example, we experimented with lightweight fabric that moved dynamically in response to the airflow generated by a set of parallel fans, providing a multisensory experience that combined visual, auditory, and haptic elements.

Furthermore, recent studies on Movement-Based Interaction have demonstrated that artefacts can be deliberately designed to move in repetitive ways, effectively communicating with people to some extent \cite{Saunderson2019}. For instance, a robotic chair that swayed back and forth enticed passers-by to sit on it \cite{Agnihotri2019}, while the speed and direction of a shaking flexible extrusion from a small box conveyed emotional expressions such as happiness and sadness \cite{Tan2016}. Building on these insights, our focus was to design visually synchronised and interpretable motions for the air filters in the WeWattTree, manipulated by the wind generated from human pedalling. This paper presents our preliminary findings on the material design and kinetic conceptualisation of the WeWattTree. Through this study, we aim to emphasise the importance of material exploration in multimodal interaction design, as well as the vast potential of integrating multiple functionalities into a single public interactive system that extends beyond traditional displays.

\section{Motion Gesturing}

\begin{figure}[b!]
    \centering
    \includegraphics[width=\linewidth]{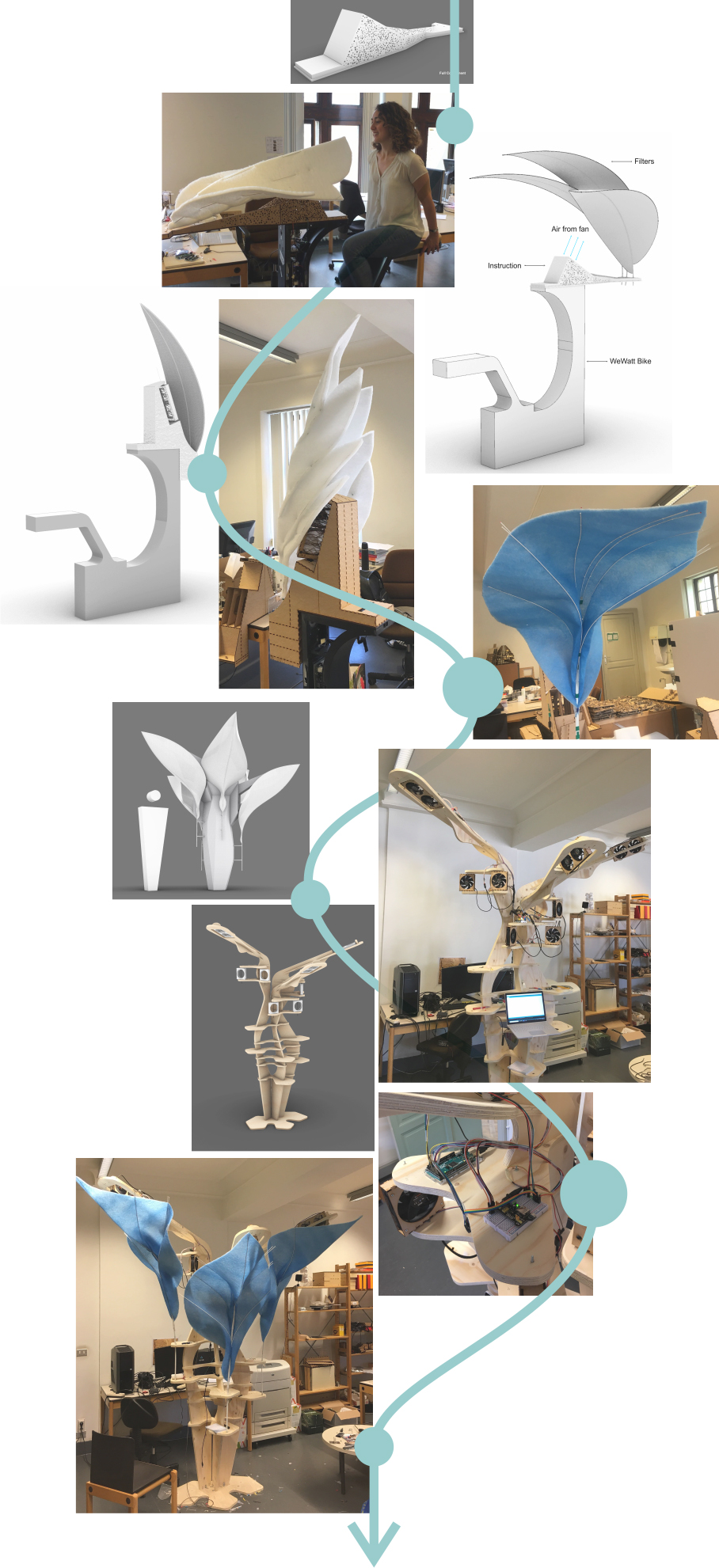}
    \caption{The design process of the WeWattTree with material exploration and iterative prototyping.}
    \label{fig:DIS2020-process}
\end{figure}

We aimed to convey several predefined meanings to people in the immediate vicinity of the WeWattTree through purposefully crafted motion gestures. For instance, we aimed to attract the attention of passers-by by generating 'random' motion gestures when the WeWattTree was not in use, encouraging them to approach and investigate the intriguing behaviour up close. Subsequently, the material manifestation of the exposed air purification filters being agitated by a series of fans, combined with the tree-like outer appearance, should help bystanders comprehend its actual functionalities, potentially inspiring some to engage in the pedalling activity. Lastly, we anticipated that the deliberate synchronisation of the motion gestures among the multiple leaves could prompt multiple bikers to synchronise their pedalling paces, which in itself requires some form of social, verbal or bodily interaction to negotiate and discuss with one another.

\subsection{Air Filtering Leaves}
To effectively convey to bystanders that the dynamic motions of the WeWattTree embodied the air purification process of a biological tree, we designed the air filters to visually resemble organic leaves. This design goal required an iterative exploration, as shown in \autoref{fig:DIS2020-process}, of the combination of physical materials and technical components to achieve an ideal shape that balances two key aspects: 1) being easily agitated, producing visually compelling and diverse motion gestures; and 2) being easily controlled to ensure consistent and robust reproducibility of the resulting gestures. To achieve this goal, we employed a combination of physical experiments and computational simulations, including airflow visualisation and kinesthetic shape optimisation. This approach enabled us to understand the dynamic repercussions of specific material properties, leading to the determination of optimal shapes and postures for the air filters.

\begin{figure}[ht]
    \centering
    \includegraphics[width=\linewidth]{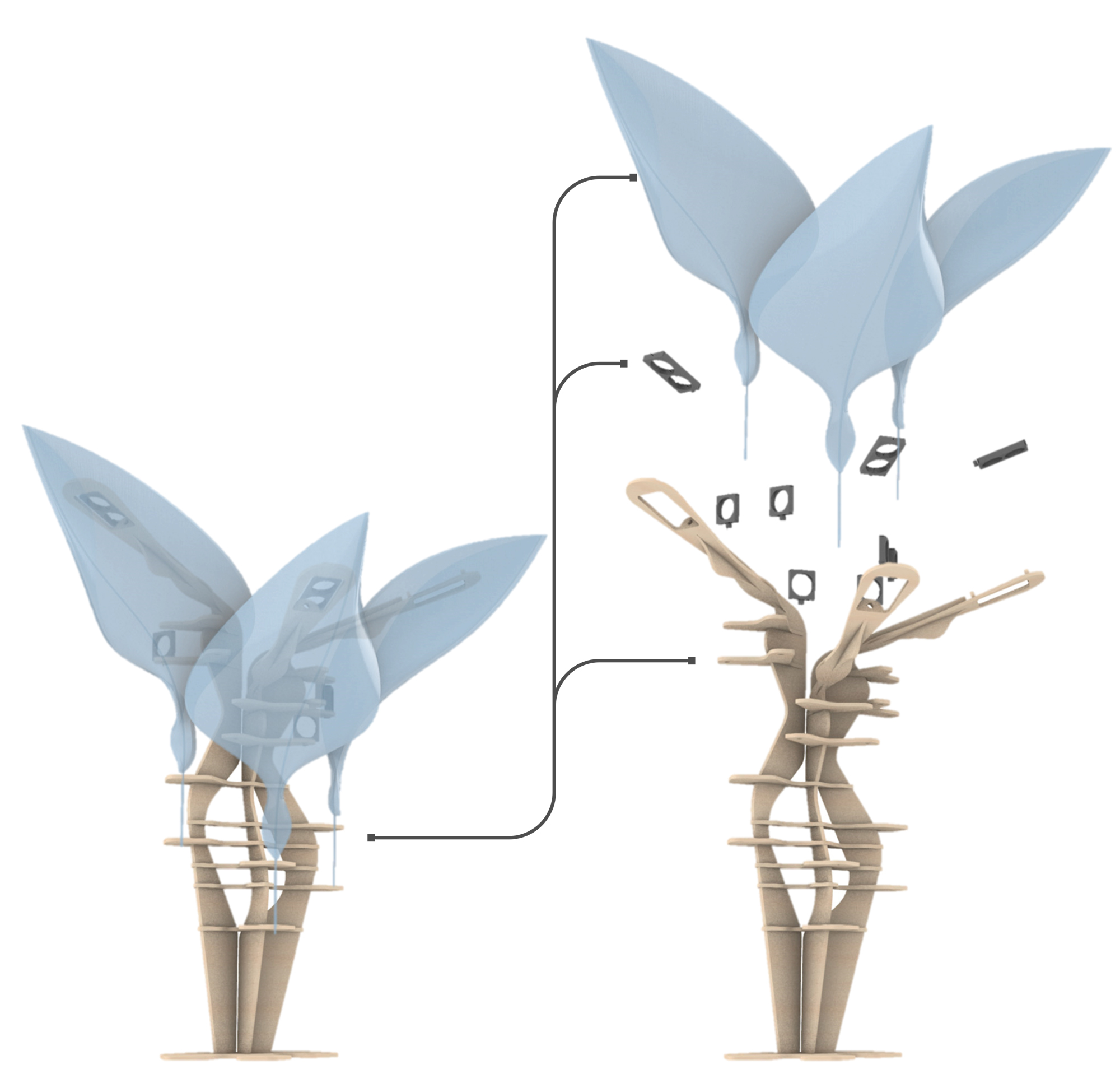}
    \caption{The components of the WeWattTree: the wooden frame, the actuators (fans and motors) and the air filtering leaves.}
    \label{fig:DIS2020-combo}
\end{figure}

\noindent The final design of the WeWattTree thus comprised three distinct air filtering leaves, each supported by a flexible skeleton. This flexibility not only created an organically curved shape highly sensitive to air flow, but also introduced a touch of randomness, resulting in slightly different motion gestures for each leaf. Indirectly actuated by a set of four electronically synchronised fans (\(1.68 W\) each), the leaves collaboratively convey a dynamic rhythm. To allow for instantaneous motion control, we connected each fan to a stepper motor that can change its physical direction, redirecting the airflow away from a leaf if needed. The total energy required to drive each leaf is \(9\) Watts, indicating that the entire installation can be powered by one biker. As shown in \autoref{fig:DIS2020-combo}, all hardware components were mounted on a digitally fabricated wooden structure, consisting of interlocking planar modules that can be disassembled for easy transportation.

\begin{figure}[ht]
    \centering
    \includegraphics[width=\linewidth]{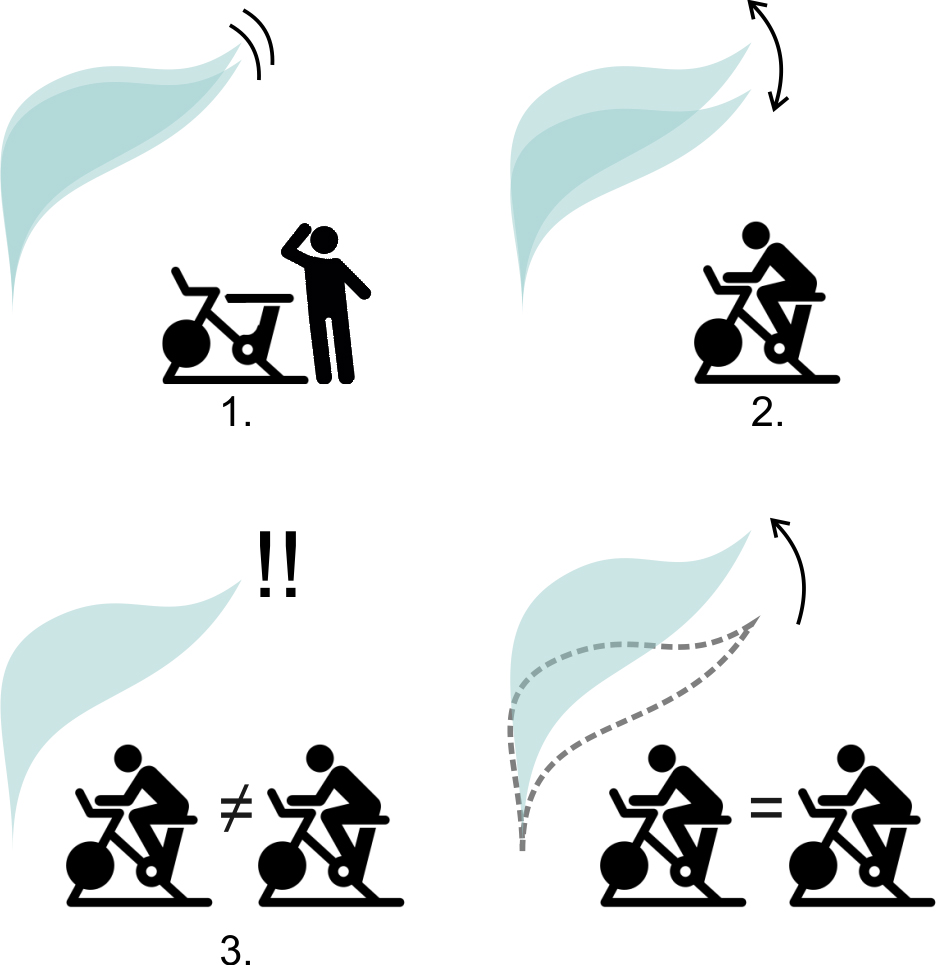}
    \caption{Example of the WeWattTree motion gestures: 1. recruitment gestures; 2. motivational gestures; 3. social gestures.}
    \label{fig:DIS2020-inter}
\end{figure}

\subsection{Motion Grammar}
From the three kinetic variables, including \textit{amplitude}, \textit{velocity}, and \textit{interval}, we developed a custom motion gesture grammar. The \textit{amplitude} represented the distance the leaf moved from its initial position to the maximum extent before reversing; \textit{velocity} denoted the average time for a leaf to complete its movement cycle, encompassing the initial position to the amplitude and back; and \textit{interval} represented the pausing time between moving cycles. A `smooth' movement is achieved when the interval value is close to \(0\) seconds. Combining these variables, the grammar generates three types of motion gestures, as depicted in \autoref{fig:DIS2020-inter}:

\begin{itemize}
    \item \textit{Recruitment gestures}: Designed to catch the attention of passers-by, these gestures involve erratic rising (big amplitude, high velocity, no interval).
    \item \textit{Motivational gestures}: Aimed at encouraging biking engagement, these gestures mimic people's cadence by mapping sensed data from the bike to the variables.
    \item \textit{Social gestures}: Intended to promote social cooperation while biking, these gestures '\textit{angrily}' interrupt the rhythm when people are out of sync (large amplitude, high velocity, long interval), or '\textit{ring up}' to reward their synchronisation (small amplitude, increasing velocity, no interval).
\end{itemize}

\section{Discussion}
During the iterative design process, the WeWattTree was tested by researchers and visitors of the research group. During these evaluation sessions, observations were made and informal interviews were conducted.

\subsection{Conceptual Meanings}
The materiality of the WeWattTree embodied the integration of multiple meanings. However, its physical affordance was not able to fully convey its combined functionalities to people. For instance, while people understood the environmental meaning of the organic shape, they did not expect it to be an air filter because it did not look industrial or functional enough. On the other hand, although the repetitive motions conveyed the filtering functionality, it appeared too rigid to be considered natural. As a result, we realised that by balancing multiple functionalities in one design, there was a risk of interference among them, potentially limiting the effective communication of meanings to people.

\subsection{Material Exploration}
As a movement-induced display in physical reality, the motions of the WeWattTree heavily depended on the combination of material and structural properties. Consequently, we experienced that the dynamic effects resulting from multiple simultaneously changing external stimuli, such as moving airflows, could not be fully simulated or modelled computationally. The manual prototyping and fine-tuning process, meanwhile, required a significant amount of time and effort, as we had to learn tacit knowledge through hands-on experimentation. Based on our experience, we suggest that an optimal workflow should involve first testing the conceptual choices in virtual simulation to validate the ideas. Afterwards, physical experimentation and fine-tuning can help explore the possibilities of motion and achieve the desired effects.

\subsection{Performance as Display} 
The WeWattTree explored the potential of a physically moving installation to convey a \textit{performance} that can be interpreted and enjoyed, but also relied on the active participation of its users. In this sense, the collaborative efforts of the bikers to synchronise their movements could become as much of a spectacular performance as the installation itself. However, the ultimate purpose of the WeWattTree goes beyond entertainment; it aims to bring strangers together and foster a sense of community while working towards a shared goal, in this case, purifying the air for everyone. This stands in stark contrast to the otherwise individual and self-centred motive of simply charging one's own phone through personal actions. By encouraging cooperation, the WeWattTree serves as a reminder of the collective impact that small, positive actions can have on the environment at large.

\section{Conclusion}
This paper presents the design and materialisation of the WeWattTree, a human-powered public display that encouraged social biking through movement-based interaction with air filtering functionality. We explored the challenges and risks of integrating multiple functionalities into a single concept, and highlighted the importance of material exploration in realising the design through physical experimentation and computational simulation. Furthermore, we discussed the potential of a physically moving display to facilitate social performances. Currently, the WeWattTree served as a working demonstrator, showcasing how human pedalling can be utilised for purposes beyond simply charging phones. Our future plans involved deploying the WeWattTree alongside three WeWatt bikes in an in-the-wild user study. Through this study, we aimed to gauge how passers-by understand and respond to the generated motion gestures to evaluate their effectiveness.

\begin{acks}
The research reported in this paper is funded by the European Union Interreg project Nano4Sports.
\end{acks}

\bibliographystyle{ACM-Reference-Format}
\balance 
\bibliography{sample-base}

\end{document}